\documentclass{moriond}
\usepackage{slashed}
\usepackage{caption}

\def\be{\begin{equation}}
\def\ee{\end{equation}}
\def\bea{\begin{eqnarray}}
\def\eea{\end{eqnarray}}

\newcommand{\diff}{\mathrm{d}}

\newcommand{\Photo}{\includegraphics[height=35mm]{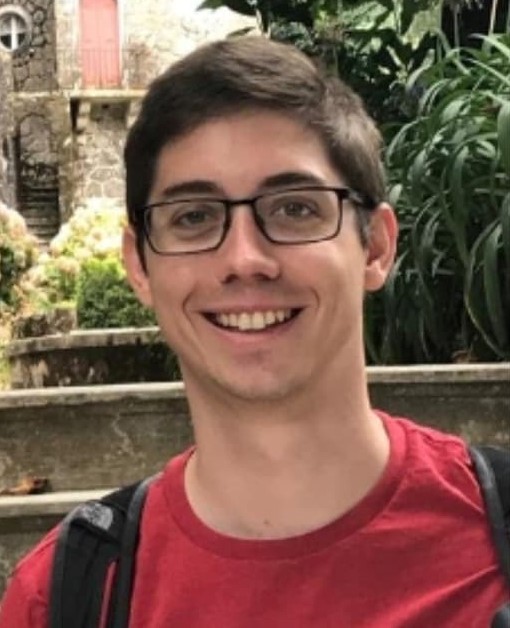}}

\begin{document}
{\hspace{12cm} ULB-TH/21-06}
\vspace*{4cm}
\title{Dark matter from dark photons}

\author{L. VANDERHEYDEN}

\address{Service de Physique Th\'eorique, Universit\'e Libre de Bruxelles,\\ Boulevard du Triomphe, CP225, 1050 Brussels, Belgium}

\maketitle\abstracts{
In this talk, based on \cite{Hambye2019}, I explained how the observed dark matter (DM) relic abundance can be accounted for in models composed of three sectors (the DM, the Standard Model (SM) and a light mediator) connected to each other. This scenario is explored in the context of the massive dark photon model in which the DM is connected to the SM through the kinetic mixing between the hypercharge gauge boson and the vector gauge boson associated to a new $U(1)'$ gauge symmetry. In such portal models, the DM relic abundance can be produced in no less than nine regimes along five dynamical mechanisms. In particular, I emphasize the sequential freeze-in dynamical mechanism which can occurs when the dark photon is massive and consists in two successive freeze-in mechanisms.}

\section{Introduction}
The idea that the DM could belong to a hidden sector (HS) is an interesting possibility, especially given the fact  that so far no experiments has come with a clear evidence for the nature of DM as a particle, even if a HS could be connected to the visible sector (VS) made of SM particles through a portal \cite{Patt2006}. How to account for the observed DM relic abundance in portal models has already been studied in various works, see in particular the detailed work of for the kinetic mixing portal case with a massless dark photon \cite{Chu2011}. We review here the massive dark photon case, as studied at length in \cite{Hambye2019}. The kinetic mixing model introduces a new $U(1)'$ gauge symmetry under which the DM candidate, a Dirac fermion $\chi$, is charged. The gauge vector boson $B'^{\mu}$ associated to this new symmetry is called dark photon and kinetically mixes with the SM massless boson $B^{\mu}$ through the parameter $\hat{\epsilon}$. The Lagrangian of this portal model can be written as following,

\begin{eqnarray}
\mathcal{L}' = -\frac{1}{4}B'^{\mu\nu}B'_{\mu\nu}-\frac{\hat{\epsilon}}{2}B^{\mu\nu}B'_{\mu\nu}+\frac{1}{2}m_{\gamma '}^{2}B'^{\mu}B'_{\mu}+i\bar{\chi}\slashed{D}\chi-m_{\rm DM}\bar{\chi}\chi,\label{eq:lag}
\end{eqnarray}

\noindent where the covariant derivative of $\chi$ is expressed in terms of the dark coupling $e'$: $D^{\mu}=\partial^{\mu}+ie'B^{\mu}$.

The massive dark photon case brings an all range of new phenomenons which do not appear in the massless case. In order to see that, let us extract, from the Lagrangian given in Eq. \ref{eq:lag}, the actual relevant interactions for DM production.

\section{Three sectors - three connectors}
\begin{figure}[h!]
\centering
\includegraphics[scale=0.42]{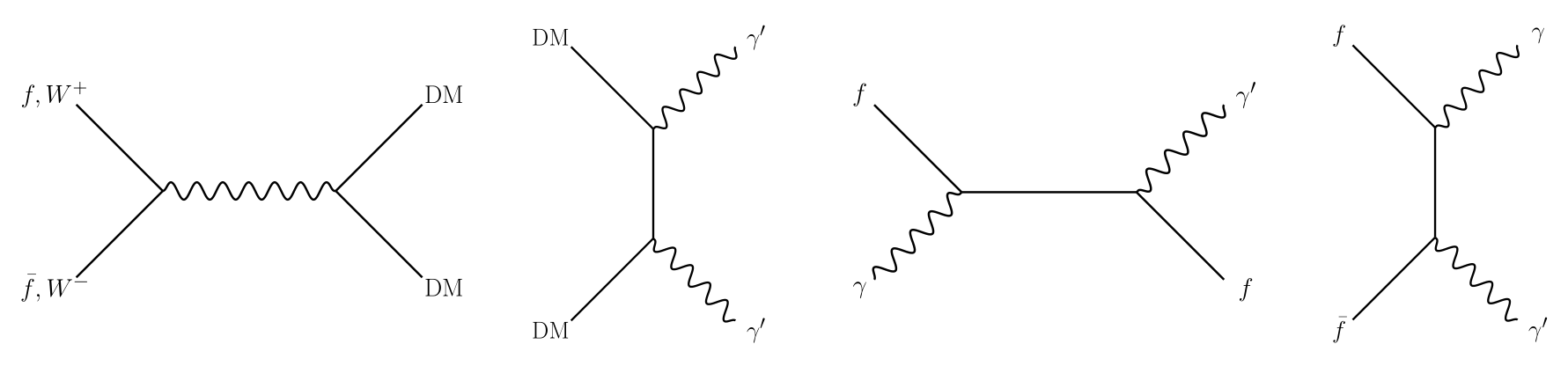}
\caption{Feynmann diagrams of the relevant processes for DM and dark photon production in the kinetic mixing portal model for the massive dark photon case.}\label{fig:diag}
\end{figure}

It is necessary to re-express the Lagrangian of Eq. \ref{eq:lag} in the basis of physical states (i.e. kinetic and mass eigenstates). To do so, one needs to perform first a non-orthogonal transformation followed by the Weinberg rotation ($\theta_{W}$) and, finally, an orthogonal diagonalisation of the remaining mass terms, see \cite{Hambye2019} for more details. At leading order in the mixing parameter $\epsilon\equiv\hat{\epsilon}\cos\theta_{W}$, one has

\begin{eqnarray}
\mathcal{L}_{I} = -\frac{e}{\sin\theta_{W} \cos\theta_{W}}J^{\mu}_{Z}Z_{\mu}-e'J^{\mu}_{\chi}\left(\epsilon \sin\theta_{W}Z_{\mu}-\gamma '_{\mu}\right)+eJ^{\mu}_{EM}\left(\gamma_{\mu}-\epsilon \cos\theta_{W}\gamma '_{\mu}\right),\label{eq:lag_I}
\end{eqnarray}

\noindent for the interaction part of the Lagrangian. One sees from Eq. \ref{eq:lag_I} that new connections arise in the massive dark photon scenario with respect to the massless case, see Figure \ref{fig:diag}. Indeed, when the dark photon is massless, one has two massless gauge bosons which can be rotated to go in a basis where the two last diagrams of Figure \ref{fig:diag} do not exist and in which one cannot produce directly the dark photon from the SM bath as well known, see e.g. \cite{Redondo2008}. Processes depicted in Figure \ref{fig:diag} are all driven by one of the three connections of the theory: $\alpha '\equiv\frac{e'^{2}}{4\pi}$ rules the interaction between the DM and dark photon particles, $\epsilon$ rules interactions between the dark photon and SM particles and $\kappa'\equiv\epsilon\sqrt{\frac{\alpha'}{\alpha}}$ rule interactions between the DM and SM particles. These three populations (the DM, mediator and SM particles) are then connected to each other through three connections instead of two as in the massless case, see Figure \ref{fig:triangles}. Notice that taking into account finite temperature corrections the $\epsilon$ coupling is effectively replaced by $\epsilon_{\rm eff}\equiv \epsilon\times m_{\gamma'}^{2}/(m_{\gamma'}^{2}-\Pi_{\gamma,T})$ where $\Pi_{\gamma,T}$ is the self-energy of the transverse propagating photons $\Pi_{\gamma,T}$,\cite{Redondo2008,An2013,Redondo2013} so that taking the limit $m_{\gamma'}\rightarrow 0$ leads well to the massless case with only 2 physical connections.

Depending on the strength of these three couplings, the three populations (the DM, mediator and SM particles) may or may not thermalise with each others, leading to DM relic abundance production mechanisms which are phenomenologically very different. The DM relic abundance depends on four independent parameters which are the DM and dark photon masses ($m_{\rm DM}$ and $m_{\gamma'}$) as well as two of the three couplings (we choose $\alpha'$ and $\kappa'$). Then, fixing the masses, we can plot contours of the DM relic abundance in the $\kappa'- \alpha'$ plane starting with an empty dark sector (i.e. assuming that no DM and no dark photon are produced at the end of inflation) then going through all production regimes all the way from phases where the DM particles never thermalise with any other particles to phases where all particles in the early Universe thermal bath thermalise with each other. This has been done in Figure \ref{fig:mesa} for $\left(m_{\rm DM},m_{\gamma'}\right)=\left(3\,{\rm GeV},1\,{\rm GeV}\right)$ in the left panel and $\left(m_{\rm DM},m_{\gamma'}\right)=\left(100\,{\rm GeV},10\,{\rm GeV}\right)$ in the right panel. 

In order to get the phase diagrams shown in left and right panels of Figure \ref{fig:mesa}, one has to integrate a set of Boltzmann equations for the DM and the mediator yields as functions of $x\equiv m_{\rm DM}/T$:
\newpage

\begin{figure}[h!]
\centering
\includegraphics[scale=1.0]{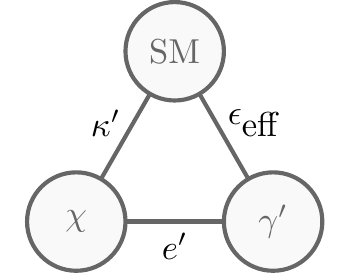}
\includegraphics[scale=1.0]{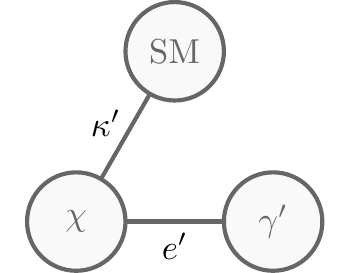}
\caption{Three sectors and their connections in the massive (left) and massless (right) dark photon cases.}\label{fig:triangles}
\end{figure}

\begin{eqnarray}
xHs\frac{\diff Y_{\rm DM}}{\diff x}&=&2\langle \sigma_{\gamma'\gamma' \rightarrow \chi\bar{\chi}} v\rangle n^2_{\gamma'}-2\langle \sigma_{\chi\bar{\chi} \rightarrow \gamma'\gamma'} v\rangle n^2_{\chi}\nonumber\\
&&\hspace{-2cm}+2\langle \sigma_{\chi\bar{\chi} \rightarrow f\bar{f}} v\rangle \left[({n_{\chi}^{\rm eq}})^2-n^2_{\chi} \right]+2\langle \Gamma^D_{Z \rightarrow \chi\bar{\chi}} \rangle\frac{n_{Z}^{\rm eq}}{(n_{\chi}^{\rm eq})^{2}} \left[{(n_{\chi}^{\rm eq}})^2-n^2_{\chi} \right]\label{eq:YDM}\\
xHs\frac{\diff Y_{\gamma'}}{\diff x}&=&2\langle \sigma_{\chi\bar{\chi} \rightarrow \gamma'\gamma'} v\rangle n^2_{\chi}-2\langle \sigma_{\gamma'\gamma' \rightarrow \chi\bar{\chi}} v\rangle n^2_{\gamma'}\nonumber\\
&&\hspace{-2cm}+\left(\langle \sigma_{f\gamma \rightarrow f\gamma'} v\rangle+\langle \sigma_{f\bar{f} \rightarrow \gamma\gamma'} v\rangle\right) \left[({n_{\gamma'}^{\rm eq}})-n_{\gamma'} \right]+2\langle \Gamma^D_{{\rm H} \rightarrow \gamma'\gamma'} \rangle\frac{n_{\rm H}^{\rm eq}}{(n_{\gamma'}^{\rm eq})^{2}} \left[{(n_{\gamma'}^{\rm eq}})^2-n^2_{\gamma'} \right]. \label{eq:Ymed}
\end{eqnarray}

\noindent where $\Gamma^D$ refers to the decay rate of the SM $Z$ boson into two DM particles or of the SM $H$ boson into two $\gamma'$. We kept a sum over all SM fermions channels implicit for simplicity.

Integrating Eqs. \ref{eq:YDM} and \ref{eq:Ymed} gives rise to no less than nine regimes along five distinct dynamical mechanisms\footnote{In Fig. 3 each corner, that the dashed line does, indicates a change of regime. In the massless mediator case, five regimes and four dynamical mechanisms were already found and discussed in \cite{Chu2011} for the Kinetic Mixing model and a scalar portal.}. This five dynamical production mechanisms which are the freeze-in (I), the sequential freeze-in (II), the reannihilation (III), the secluded freeze-out (IV) and, finally, the freeze-out (V) are distinguishable looking at which of the three connections of left panel of Figure \ref{fig:triangles} thermalises or not as well as looking at which connection dominates over the others.

\begin{figure}[h!]
\centering
\includegraphics[scale=0.595]{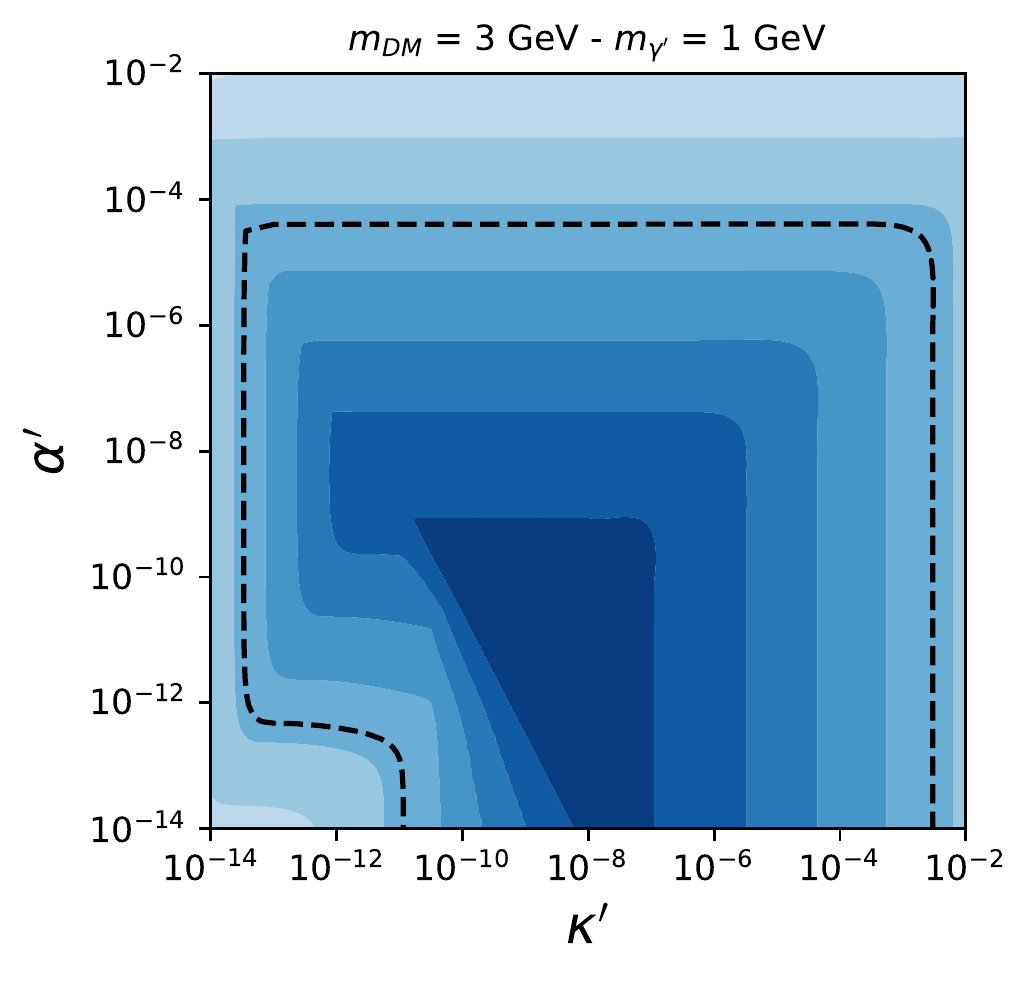}
\includegraphics[scale=0.595]{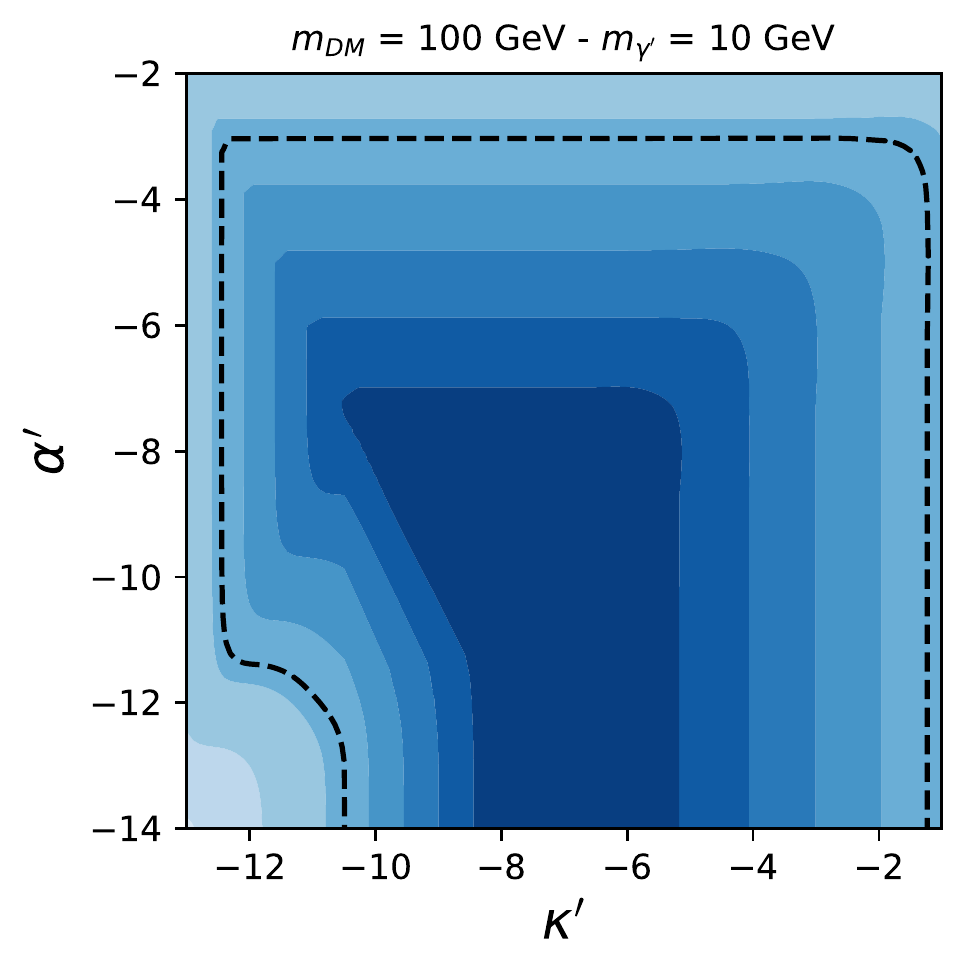}
\caption{Contours of the DM relic density in the $\alpha'-\kappa'$ plane for $\left(m_{\rm DM},m_{\gamma'}\right)=\left(3\,{\rm GeV},1\,{\rm GeV}\right)$ (left panel) and $\left(m_{\rm DM},m_{\gamma'}\right)=\left(100\,{\rm GeV},10\,{\rm GeV}\right)$ (right panel) in the kinetic mixing portal model for the massive dark photon case. The dashed contour indicates $\Omega_{\rm DM}h^{2}=0.1188$.}\label{fig:mesa}
\end{figure}

As an example, let us explain in more details the Sequential Freeze-In regime (II) which corresponds to the vertical regime on the extreme left of each panel of Figure \ref{fig:mesa}. In this part of the parameter space, one can check that none of the three sectors thermalises with any of the two others. That is to say that, 

\begin{eqnarray}
\Gamma_{\rm SM \leftrightarrow DM}<H\hspace{0.5cm} {\rm \&} \hspace{0.5cm}\Gamma_{{\rm SM} \leftrightarrow \gamma'}<H\hspace{0.5cm} {\rm \&} \hspace{0.5cm}\Gamma_{{\rm DM} \leftrightarrow \gamma'}<H,
\end{eqnarray}

\noindent where $\Gamma_{\rm X\leftrightarrow Y}$ stems for the total interaction rate between sector $X$ and sector $Y$ and with $H$ the Hubble expansion rate. Thus, as no sector thermalises with any another one and starting with an empty dark sector,\footnote{The opposite option (more initial condition dependent and which we do not study here) would be that the final amount of DM would be mainly due to the presence of many HS particles at the end of inflation with little role of the portal.} the DM relic abundance is slowly produced by slow out-of-equilibrium processes from SM. Along this regime the value of the DM-to-SM coupling, $\kappa'$, turns out to be too low to directly play this role. Thus DM cannot be produced from an usual $SM\rightarrow DM$ freeze-in regime (first diagram of Figure \ref{fig:diag}). Nevertheless, it turns out along this regime that the SM can slowly produces enough dark photon through out-of-equilibrium ${\rm SM}\rightarrow\gamma'$ processes, for these dark photons to slowly produce subsequently enough DM through $\gamma' \rightarrow$ DM processes, see the two last diagrams depicted in Figure \ref{fig:diag}. In other words these dark photons never enter thermal equilibrium, but can still slowly produce DM through out-of-equilibrium processes $\gamma'\rightarrow{\rm DM}$ (second diagram of Figure \ref{fig:diag}). Therefore, it is possible that the whole DM observed today can be produced through these two successive freeze-in mechanisms, a "Sequential Freeze-In".

\section{Summary}
The DM relic abundance in a three sectors-three connections scenario can be set through five different dynamical mechanisms, and this along nine distinct regimes. We illustrated these results through the massive dark photon model of Eq. \ref{eq:lag}, generalizing the massless dark photon case \cite{Chu2011}. Among the four new dynamical mechanisms, a new DM production regime which we dubbed "Sequential Freeze-In" (II) consists first of a slow out-of-equilibrium production of dark photons from SM particles followed by a slow out-of-equilibrium production of DM particles from dark photons particles. The other new regimes are DM freeze-in production from a thermalised population of dark photons (Ib), reannihilation (IIIa) and secluded freeze-out (IVa) through the freeze-in production of dark photons, see \cite{Hambye2019}. Although we considered the kinetic mixing portal model to illustrate our findings, we argue that the behaviours we presented are actually more generics. They are characteristics of DM scenarios in which the DM is connected to the SM bath through a portal carried by a new particle which would give rise to three sectors connected to each others by three connections due to two different couplings, here $\alpha'$ and $\epsilon$. This also holds in particular for Higgs portal models, see \cite{Hambye2019} and also \cite{Belanger2020}.

\section*{Acknowledgments}
I thank the organizers for this special edition of the Rencontres de Moriond. I also thank T. Hambye, M.H.G. Tytgat and J. Vandecasteele for collaboration on the work discussed in this note. My work is supported by the FRIA grant from FRS-FNRS.

%


\section*{References}

\begin{thebibliography}{99}
\bibitem{Hambye2019}T. Hambye, M. H. G. Tytgat, J. Vandecasteele, and L. Vanderheyden, Phys. Rev. D 100,
095018 (2019)
\bibitem{Patt2006}B. Patt and F. Wilczek, arXiv:hep-ph/0605188.
\bibitem{Chu2011}X. Chu, T. Hambye, and M. H. G. Tytgat, J. Cosmol. Astropart. Phys. 05 (2012) 034
\bibitem{Redondo2008}J. Redondo, J. Cosmol. Astropart. Phys. 07 (2008) 008
\bibitem{An2013}H. An, M. Pospelov, and J. Pradler, Phys. Lett. B 725, 190 (2013)
\bibitem{Redondo2013}J. Redondo and G. Raffelt, J. Cosmol. Astropart. Phys. 08 (2013) 034
\bibitem{Belanger2020}G. Bélanger, C. Delaunay, A. Pukhov and B. Zaldivar, Phys. Rev. D 102, 035017 (2020)

%
%
%

\end{thebibliography}
\end{document}